\documentclass[twoside, 10pt]{article}
\usepackage{zif,epsfig}

\begin{document}
\setlength{\textheight}{7.05truein}    %FOR 2ND PAGE ONWARDS

\runninghead{Algebra, Geometry and Time
Encoding of Quantum Information}
            {Michel Planat, Metod Saniga}

\normalsize\textlineskip \thispagestyle{empty}
\setcounter{page}{409}

%\copyrightheading{Vol.}{No.}{Year}{Page Nos.}
\copyrightheading{}{}{2005}{409--426}

\vspace*{0.88truein}

\alphfootnote

\fpage{409}

\centerline{\bf
%%%%%%%%%%%%%%%%%%%%%
%Put in titiles here
%%%%%%%%%%%%%%%%%%%%%
ABSTRACT ALGEBRA, PROJECTIVE GEOMETRY AND} \vspace*{0.035truein}
\centerline{\bf TIME ENCODING OF QUANTUM INFORMATION}
\vspace*{0.37truein} \centerline{\normalsize
%%%%%%%%%%%%%%%%%%%%%%%%%%%%%%%%%%%%
%put authors' name and address here
%%%%%%%%%%%%%%%%%%%%%%%%%%%%%%%%%%%%
MICHEL PLANAT} \vspace*{0.015truein} \centerline{\small\it
FEMTO-ST, University of Franche-Comt\'{e}, 32 Avenue de
l'Observatoire} \baselineskip=11pt \centerline{\small\it 25044,
Besan\c{c}on, France} \baselineskip=11pt \centerline{\small\it
(planat@lpmo.edu)}

\vspace*{11pt} \centerline{\normalsize METOD SANIGA}
\vspace*{0.015truein} \centerline{\small\it Astronomical
Institute, Slovak Academy of Sciences} \baselineskip=11pt
\centerline{\small\it SK-05960 Tatransk\'{a} Lomnica, Slovak
Republic} \baselineskip=11pt \centerline{\small\it
(msaniga@astro.sk)}

\vspace*{0.225truein}
%\publisher{(received date)}{(revised date)}

\vspace*{0.22truein}

%% \abstracts{first paragraph}{second paragraph}{third paragraph}
%% If there is only one paragraph, just keep the second and third empty
%% like the following one
\abstracts{{\bf Abstract:} Algebraic geometrical concepts are
playing an increasing role in quantum applications such as coding,
cryptography, tomography and computing. We point out here the
prominent role played by Galois fields viewed as cyclotomic
extensions of the integers modulo a prime characteristic $p$. They
can be used to generate efficient cyclic encoding, for
transmitting secrete quantum keys, for quantum state recovery and
for error correction in quantum computing. Finite projective
planes and their generalization are the geometric counterpart to
cyclotomic concepts, their coordinatization involves Galois
fields, and they have been used repetitively for enciphering and
coding. Finally, the characters over Galois fields are fundamental
for generating complete sets of mutually unbiased bases, a generic
concept of quantum information processing and quantum
entanglement. Gauss sums over Galois fields ensure minimum
uncertainty under such protocols.  Some Galois rings which are
cyclotomic extensions of the integers modulo $4$ are also becoming
fashionable for their role in time encoding and mutual
unbiasedness.}{}{} \vspace*{10pt}

%\vspace*{3pt}
%\communicate{to be filled by the Editorial}

\vspace*{1pt}\textlineskip    %) USE THIS MEASUREMENT WHEN THERE IS
   %) A SECTION HEADING
%\vspace*{-0.5pt}
%\noindent
%%%%%%%%%%%%%%%%%%%%%%%%%%%%%%%%
%put the text of the paper here
%%%%%%%%%%%%%%%%%%%%%%%%%%%%%%%%
\section{Introduction}
Many objects of our today life would not have been designed
without the revolution of knowledge undertaken one century ago:
quantum mechanics. But many philosophers, as well as scientists, are
still not satisfied with its abstract interpretation of the
physical world. The operational formalism of quantum mechanics can
answer almost every question about the observable quantities, but
we would like to know more about the quantum machine. We had time
and space in the old continuous machinery of the nineteenth
century physics; where do they
reside now? According to the Heisenberg indeterminacy
principle, there are gaps in our time description of quantum
processes that we cannot fill: accuracy in the isolation of time
events means a lack of knowledge of their energy. The same for
position in space of a particle which is complementary to the
momentum. Some scholars are convinced that we, as humans, are
partly responsible for tiny impacts such a particle may suffer
during an experiment. The loose of realism would be inherent to
the realm of quantum mechanics.

Let us point out that more knowledge about quantum processes may
be obtained thanks to  quantum information theory --- the
recent marriage of quantum mechanics and information theory. In
the last decade new concepts with quantum bits (qubits), such as
qu-cryptography, qu-teleportation, qu-cloning, qu-computing and
qu-money have been implemented \cite{BouwmeesterOO}. They have
grown upon a big stone erected in 1935: the EPR paradox about the
entanglement of quantum states. Qubit entanglement (and its
generalization to qudits, i.e. many-level quantum states)
is the main resource of the newly emerged quantum information technology.

The goal of this paper is to revisit some of the objects of
quantum information theory using finite algebraic geometrical
concepts such as finite fields (also known as Galois fields), and
to give them a geometrical setting. In doing so, a kind of discrete space-time
emerges, time being connected to algebraic ideals and space
to finite geometries. The notion of a character
maps elements of the Galois fields to the quantum states of
interest.

\section{Time and Its Relation to Ideals}

\subsection{Ideals and the residue class ring}

Let us start our quest of the nature of time in the algebraic
world. Our objects are elements of a finite set which is a ring,
$\mathcal{R}$, i.e. the set endowed with two operations ``$+$" and
\lq\lq$\cdot$". The ring is a group with respect to addition; the
product of two elements is in $\mathcal{R}$ and it is both associative and
distributive with respect to addition. One needs the concept of an
ideal $\mathcal{I}$ in $\mathcal{R}$, denoted
$\mathcal{I}\triangleleft\mathcal{R}$, which is a subset of
$\mathcal{R}$ such that $\forall a \in \mathcal{I}$, $\forall r
\in \mathcal{R}$ one has both $ar \in \mathcal{I}$ and $ra \in
\mathcal{I}$. In other words, with the concept of an ideal one pins
each element of $\mathcal{R}$ into the subset $\mathcal{I}$; and
with the concept of a principal ideal $\mathcal{I}=(a)$, a single
element $a$ generates the whole ideal. For a commutative ring
$\mathcal{R}$ with an identity the definition is: $(a)=a
\mathcal{R}=\{ar,~r \in \mathcal{R}\}$. A familiar example is the
ring of integers $\mathcal{R}=\mathcal{Z}=\{\cdots,-2,-1,0,+1,+2,\cdots \}$.
The principal ideal generated by the number $a=3$ in
$\mathcal{Z}$ is $(a)=\{\cdots,-6,-3,0,3,6,\cdots\}$.

The next important object is the concept of a residue class of $a$
modulo $\mathcal{I}$, which consists of all elements $[a]=\{a+c,~
\forall c\in \mathcal{I}\}$ and is useful to partition the ring
into disjoint classes (or cosets). The set of classes has the
property to be a ring, called the residue class ring and denoted
$\mathcal{R}/\mathcal{I}$. For the integers $\mathcal{Z}$ modulo
the ideal $(3)$, one gets the three classes $[0]=0+(3)$,
$[1]=1+(3)$ and $[2]=2+(3)$, and the residue class ring is
 $\mathcal{Z}/(3)=F_3$, where $F_3$ is the unique field with $3$
 elements. It is known that for a prime number $p$, $Z/(p)=\mathcal{Z}_p=F_p$,
 where $\mathcal{Z}_p$ is the set of integers modulo $p$ and $F_p$ the field with
  $p$ elements. But, for example, $Z/(4)$ is not a  field since $2.2=4=0$ and thus $2$ divides $0$.

\subsection{Polynomial rings, Galois fields and their representations}
\label{Galois}

Let us now consider a ring  $\mathcal{R}[x]$ of polynomials with
coefficients in  $\mathcal{R}$
\begin{equation}
\mathcal{R}[x]=\{a_0+a_1x+\cdots+a_nx^n\},~~a_i\in \mathcal{R}.
\label{eqn1}
\end{equation}
One says that $g \in \mathcal{R}[x]$ is irreducible if it cannot
be factored in $\mathcal{R}$; e.g. $x^2-2 \in \mathcal{Q}[x]$ is
irreducible in the field $\mathcal{Q}$ of rational numbers, but
$x^2-2=(x+\sqrt{2})(x-\sqrt{2})$ over the real numbers
$\Re$. There is an important theorem that for any
polynomial $g \in \mathcal{R}[x]$, the residue class ring
$\mathcal{R}[x]/(g)$ is a field if and only if (iff) $g$ is
irreducible over $\mathcal{R}~$\cite{Lidl83}. For example for
$\mathcal{R}=F_2=\{0,1\}$, the field with two elements, and since
$g=x^2+x+1$ is irreducible over $F_2$, then $F_4=F_2[x]/(g)$ is
the Galois field with $4$ elements $[0]=(g), [1], [x]$ and
$[x+1]$. For example
$[x]+[x+1]=x+(g)+x+1+(g)=2x+1+(g)+(g)=1+(g)=[1]$. Similarly
$[x][x]=(x+(g))(x+(g))=x^2+(g)(2x+1)=x^2+(g)=x^2-(x^2+x+1)+(g)=-(x+1)+(g)=(x+1)+(g)=[x+1]$.

It can be shown that a Galois field with $q$ elements exists iff
$q=p^m$, a power of a prime number $p$. Actually, there are several
representations of Galois fields. The first one is as a polynomial
as in (\ref{eqn1}). The second one consists of identifying the
Galois field $F_q$, with $q=p^m$, to the vector space $F_p^m$ build
from the coefficients of the polynomial. The third one uses the
property that $F_q^*=F_q-\{0\}$ is a multiplicative cyclic group.
One needs the concept of a primitive polynomial. A (monic)
primitive polynomial, of degree $m$, in the ring $F_q[x]$ is
irreducible over $F_q$ and has a root $\alpha \in F_{q^m}$ that
generates the multiplicative group of $F_{q^m}$. A polynomial
$g\in F_q[x]$ of degree $m$ is primitive iff $g(0)\neq 0$ and
divides $x^r-1$, with $r=q^m-1$.

For example, $F_8$ can be build from $\mathcal{R}=F_2$ and
$g=x^3+x+1$ which is primitive over $F_2$. One gets
$F_8=F_2[x]/(g)=\{0,1,\alpha,\alpha^2,\alpha^3=1+\alpha,
\alpha^4=\alpha+\alpha^2,\alpha^5=1+\alpha+\alpha^2,\alpha^6=1+\alpha^2\}$,
see Table 1.
\begin{table}[htbp]
\tcaption{Representations of the elements of the Galois field
$GF(8)$. }
\begin{center}
\begin{tabular}{|c|c|c|}
%\begin{ruledtabular}
\hline
 as powers of $\alpha$ &as polynomials&as $3$-tuples in $\mathcal{Z}_2^3$\\
\hline \hline
0 & 0 &$(0,0,0)$\\
\hline
1&1&$(0,0,1)$\\
\hline
$\alpha$ & $\alpha$ & $(0,1,0)$\\
\hline
$\alpha^2$&$\alpha^2$&$(1,0,0)$\\
\hline
$\alpha^3$&$1+\alpha$&$(0,1,1)$\\
\hline
$\alpha^4$&$\alpha+\alpha^2$&$(1,1,0)$\\
\hline
$\alpha^5$&1+$\alpha$+$\alpha^2$&$(1,1,1)$\\
\hline
$\alpha^6$&$1+\alpha^2$&$(1,0,1)$\\
\hline
%\end{ruledtabular}
\end{tabular}
\end{center}
\end{table}
\subsection{Cyclic codes as ideals}
\label{codes}

In one of our recent papers arithmetical functions were considered
relevant models of time evolution \cite{Planat01}. For instance,
the function $a(n)$ defined as $1$ if $n=p^m$, $p$ a prime number,
and $0$ otherwise, was found to play an important role in the study of phase
fluctuations in an oscillator. Three generic functions are met in
elementary analytical number theory. One is the Mangoldt function,
closely related to the above defined $a(n)$ function. The second
is the Euler (totient) function $\phi(n)$, which counts the number of
irreducible fractions $l/n$, ${\rm gcd}(l,n)=1$.  If one knows the
decomposition $n=\prod_i p_i^{m_i}$ as a product of prime powers,
then $\phi(n)=n\prod_i \left(1-1/p_i \right)$. The third one, the M\"{o}bius
function, codes the distribution of primes as $\mu(1)=0$,
$\mu(n)=0$ if $n$ contains a square and $(-1)^k$ if $n$ is the
product of $k$ distinct prime numbers. The last two functions
still appear in the theory of cyclic codes, as it will be illustrated
below.

In Sect.\,\ref{Galois} we defined Galois fields as the residue
class ring over a ground field $F_p$ of characteristic $p$,
generated by a polynomial $g(x)$ irreducible over $F_p$. One can
generalize this view by considering $F_q$, $q=p^n$, as the ground
field and by defining an ideal $(g)$ from a polynomial $g$ which
is irreducible over the polynomial field $F_q[x]\equiv F_q^n$.
This definition encompasses all linear cyclic codes. A linear code
is any vector subspace of $F_q^n$, and it is cyclic if one goes from one
line to the other of the generating matrix by a shift of its
elements.

All cyclic codes are constructed by all the divisors $g$ in
$F_q[x]$ of the polynomial $x^n-1$. The divisors are $g=Q_d$, the
so-called $d^{\rm{th}}$ cyclotomic polynomials, their degree is
$\phi(d)$ and they are defined as
\begin{equation}
Q_d=\prod_{d|n}(x^d-1)^{\mu(n/d)}. \label{eqn2}
\end{equation}
\label{cyclotomic}
The Mangoldt function \cite{Planat01} is a way of encoding the primes.\footnote{The Mangoldt
function, $\Lambda(n)$, plays a prominent role in the (still
unsolved) Riemann hypothesis. $\Lambda(n)$ equals $\ln (p)$ if $n=p^m$  and $0$
otherwise. Its average value oscillates around $1$ and the error
term explicitly relies on the pole at $s=1$ of the Riemann zeta
function $\zeta(s)=\sum_{\Re(s)>1} n^{-s}$, $\Re(s)>1$, on the
trivial zeros at $s=-2l$, $l>0$, of the extended zeta function
$\xi(s)=\pi^{-s/2}\Gamma(s/2)\zeta(s)$, $\Gamma (s)$ being the Gamma
function, and on the Riemann zeros presumably all located on the
critical axis $\Re(s)=1/2$.} ~In the new context of cyclic codes,
the cyclotomic polynomial also encodes the irregularity of primes.
There exists a \lq\lq zeta" function and a \lq\lq Riemann hypothesis"
for $F_q$; the latter was proved by Weil in 1948
\cite{Roquette03}.

Let us describe a linear $[n,k]$ code from its generator matrix.
We use the polynomial
\begin{equation}
g=g_0+g_1x+\cdots+g_mx^m ~\in
F_q[x]=F_q^n,~~g|(x^n-1),~~ \deg(g)=m<n.
\end{equation}
The generator matrix is as follows
\begin{equation}
 \left [\begin{array}{ccccccc} g_0 & g_1&\cdots&g_m&0&\cdots&0\\ 0 & g_0&\cdots&g_{m-1}&g_m&\cdots&0\\
 \cdots&\cdots&\cdots&\cdots&\cdots&\cdots&\cdots\\0&0&\cdots&g_0&g_1&\cdots&g_m\nonumber \\
\end{array}\right]=\left [\begin{array}{c} g \\ xg\\
 \cdots\\x^{k-1}g\nonumber \\
\end{array}\right].
\label{table1}
\end{equation}
As an example, we mention the binary Hamming code of length $n=7$,
which is obtained from $g=x^3+x+1$ of coefficients over $F_2$ and contains
$4$ elements which are the lines of the following generating matrix
\begin{equation}
 \left [\begin{array}{ccccccc} 1 & 1&0&1&0&0&0\\ 0 & 1&1&0&1&0&0\\
 0&0&1&1&0&1&0\\0&0&0&1&1&0&1\nonumber \\
\end{array}\right].
\end{equation}
The index $n$ plays the role of time and the code is thus a
$2$-valued time encoded by the cyclotomic polynomial
(2).

\section{Quantum States and Their Relation to Additive Characters}

\subsection{The additive characters}
\label{characters}

 A character $\kappa(g)$ over an abelian group
$G$ is a (continuous) map from $G$ to the field of complex numbers
$\mathcal{C}$ that is of modulus $1$, i.e. such that
$|\kappa(g)|=1$, $g\in G$.
Since there are two operations \lq\lq$+$" and \lq\lq$\cdot$" in the
field $F_q$, one can define two kinds of characters.
Multiplicative characters $\psi_k(n)=\exp(\frac{2i\pi n k}{q})$,
$k=0,\ldots,q-1$, are well known since they constitute the basis for the
ordinary discrete Fourier transform. But additive characters
introduced below are the ones which are useful to relate to
quantum information.
One first defines a map from the extended field $F_q$, $q$=$p^m$, to the
ground field $F_p$ which is called the trace function
\begin{equation}
tr(x)=x+x^p+\cdots+x^{p^{m-1}}\in F_p,~~ \forall ~x\in F_q.
\label{trace0}
\end{equation}
In addition to its property of mapping an element of $F_q$ into
$F_p$, the trace function has the following properties: $tr(x+y)=tr(x)+
tr(y),~x,y \in F_q$; $tr(ax)=a~tr(x),~x \in F_q,~a \in F_p$;
$tr(a)=ma, ~a\in F_p$; and $tr(x^q)=tr(x),~ x\in F_q.$
Using (\ref{trace0}), an additive character over $F_q$ is defined
as
\begin{equation}
\kappa(x)=\omega_p^{tr(x)},~~\omega_p=\exp \left(\frac{2i\pi}{p}\right),~~x \in F_q.
\end{equation}
It satisfies the following relation: $\kappa(x+y)=\kappa(x)\kappa(y), ~x,y \in F_q$.

\subsection{Quantum states: qubits and qudits}

Well before the development of quantum information theory
physicists developed an efficient formalism for working out
quantum states. This formalism was born (with Dirac) in the context of the
second quantization of a harmonic oscillator. The language of
kets $|u\rangle$ and bras $\langle u|$, for $u$ an element of a
Hilbert space $\mathcal{H}$, a vector space over the complex
numbers $\mathcal{C}$ equipped with a complex-valued inner product
$\mathcal{H}\times\mathcal{H} \rightarrow \mathcal{C}$, is still
in use today.

Physically, a qubit is an element of a Hilbert space
of dimension $2$, $\mathcal{H}_2$; it can represent a spin $1/2$, a
two-level atomic system, a two-polarisation state, {\it etc.} The most
general form of a qubit $|\psi \rangle$ is
\begin{equation}
|\psi \rangle=a|0\rangle +b |1\rangle,~~|a|^2+|b|^2=1~~a,b \in
\mathcal{C}.
\end{equation}
In the computational frame of a qubit base $B_0=(|0\rangle,
|1\rangle)$, we have $|0\rangle=(1,0)$ and $|1\rangle=(0,1)$. The
geometry of the qubit is the Bloch sphere \cite{Nielsen00}, with
the qubit $|0\rangle$ at the north pole and the qubit $|1\rangle$
at the south one. In what follows we will be interested in qudits, quantum
states in a generic, $q$-dimensional
Hilbert space $\mathcal{H}_q$ defined as $|\psi \rangle =\sum_{k=0}^{q-1} a_k |k
\rangle$, $\sum_k|a_k|^2=1$, $a_k \in \mathcal{C}$, although recently the particular
cases of $q$=2, 4 and 8 received a lot of attention due to their intimate link to
Hopf fibrations (see, e.g. \cite{Bernevig03}).

Another important concept for quantum measurements has recently emerged,
the one of a complete set of mutually unbiased bases (MUBs).
Besides the concept of an additive character of the Galois field
$F_q$, MUBs reveal a connection between $F_q$ and the structure of
Hilbert space $\mathcal{H}_q$. Orthogonal bases of a Hilbert space
$\mathcal{H}_q$ of finite dimension $q$ are mutually unbiased if
inner products between all possible pairs of vectors of distinct
bases equal $1/\sqrt{q}$. They are also said to be maximally non-commutative
in the sense that a measurement over one basis leaves
one completely uncertain as to the outcome of a measurement
performed over a basis unbiased to the first. For $q$=2, the eigenvectors of
ordinary Pauli spin matrices provide the
best-known example.

With a complete set of $q+1$ mutually unbiased measurements one
can ascertain the density matrix of an ensemble of unknown quantum
$q$-states, so that a natural question emerges as which
mathematics may provide the construction. It is known that in
dimension $q=p^m$  the complete
sets of mutually unbiased bases (MUBs) result from Fourier
analysis over the Galois field $F_q$ ($p$ odd)
\cite{Wootters89} or a Galois ring $R_{4^m}$($p$ even) \cite{Klapp03}.

Constructions of MUBs in odd characteristic\footnote{Fourier
analysis and MUBs in even characteristic are studied in Sect.\,\ref{Even}.} ~are related to the character sums with polynomial
arguments $f(x)$, called Weil sums,
\begin{equation}
\sum_{x \in F_q}\kappa(f(x)). \label{Weilsums}
\end{equation}
In particular (see theorem 5.38 in \cite{Lidl83}), for a polynomial
$f(x) \in F_q[x]$ of degree $d\ge 1$, with $\rm{gcd}$$(d,q)=1$, one gets
$|\sum_{x \in F_q}\kappa(f(x))|\le(d-1)q^{1/2}$.
The complete sets of MUBs are obtained as \cite{Klapp03,Planat05}
\begin{equation}
|\theta_b^a\rangle=\frac{1}{\sqrt{q}}\sum_{n\in
F_q}\psi_k(n)\kappa(an^2+bn)|n\rangle,~~a,b \in F_q, \label{MUB1}
\end{equation}
with $\psi_k(n)$ and  $\kappa (x)$  defined in Sect.\,\ref{characters}. Eq.\,(\ref{MUB1})
defines a set of $q$ bases (with index $a$) of $q$
vectors (with index $b$). Using Weil sums (\ref{Weilsums}) it is
easily shown that for $q$ odd the
bases are orthogonal and mutually unbiased to each other and to
the computational base
$\{|0\rangle,|1\rangle,\cdots,|q-1\rangle\}$ as well.

\subsection{Mutually unbiased bases as quantum phase states}
\label{MUBodd}

Dirac was the first to attempt a definition of a phase operator by
means of an operator amplitude and phase decomposition. In this
description the number operator $N$ and the phase operator
$\Theta$ are canonically conjugate such that $[N,\Theta]=i$, where
$[ ~]$ are the commutator brackets, and this equation leads to a
number--phase uncertainty relation $\delta N \delta \phi \ge 1/2$.
Quantum phase states reaching the bound are coherent, or
squeezed states. But there is a big problem in defining such a
Hermitian quantum phase operator \cite{Lynch95} using the familiar
Fock states of the quantized electromagnetic field.

One way to circumvent the problem is the use of a discrete Hilbert
space $\mathcal{H}_q$. It was shown \cite{Pegg89} that states
(\ref{MUB1}) obtained from a trivial character $\kappa_0=1$  are
eigenstates of the Hermitian phase operator
\begin{equation}
\Theta=\sum_{k\in \mathcal{Z}_q}\theta_k|\theta_k\rangle \langle
\theta_k|, \label{P&B}
\end{equation}
with eigenvalues $\theta_k=\theta_0+\frac{2\pi k}{q}$, $\theta_0$
being an arbitrary initial phase. More generally, the MUB states are
eigenstates of a \lq\lq Galois" quantum phase operator
\cite{Planat05}
\begin{equation}
\Theta_{\rm{Gal}}=\sum_{b\in F_q}\theta_b|\theta_b^a\rangle
\langle \theta_b^a|,~~a,b \in F_q, \label{GalMUB}
\end{equation}
with eigenvalues  $\theta_b=\frac{2\pi b}{q}$. The operator can be
made more explicit when combined with Eq. (\ref{MUB1}),
\begin{equation}
\Theta_{\rm{Gal}}=\frac{2\pi}{q^2}\sum_{m,n \in
F_q}\psi_k(n-m)\omega_p^{tr[a(n^2-m^2)]} S(n,m) |n\rangle \langle
m|, \label{GalMUB1}
\end{equation}
with $S(n,m)=\sum_{b\in F_q}b\omega_p^{tr[b(n-m)]}$. The
diagonal matrix elements feature the sums $S(n,n)=q(q-1)/2$, while for
the non-diagonal ones one gets
$S(m,n)=\frac{q}{1-\omega_p^{tr(m-n)}}$.

\section{Phase Fluctuations: From Ramanujan to Gauss Sums}
In the previous work \cite{PLA03}, the near classical regime of a phase-locked oscillator
has been studied and its phase fluctuations  have been related to
the irregularity of the distribution of prime
numbers. A quantum model of phase-locking was derived based on
operator (\ref{P&B}) with an additional assumption that only
elements $|\theta'_k\rangle$ with $k$ coprime to $q$ were taken into account.
As a result, the quantum phase-locking operator
\begin{equation}
\Theta_{\rm{lock}}=\sum_k\theta'_k |\theta'_k\rangle\langle
\theta'_k|\label{Qlock}
\end{equation}
can be evaluated explicitly as
\begin{equation}
\Theta_{\rm{lock}}= \frac{1}{q}\sum_{n,l}c_q(n-l)|n\rangle\langle
l|, \label{projector}
\end{equation}
where $n,l$ range from $0$ to $\phi(q)$,
$\phi(q)$ being the Euler totient function. The coefficients in the last equation are the so-called
Ramanujan sums,
\begin{equation}
c_q(n)=\sum_{{\rm gcd}(p,q)=1} \exp\left(2i\pi \frac{p}{q}
n\right)=\frac{\mu(q_1)\phi(q)}{\phi(q_1)},~~q_1=q/{\rm gcd}(q,n),
\label{Rsums}
\end{equation}
where $\mu(q)$ stands for the above-introduced M\"obius function.

For the evaluation of phase variability of states we considered a
pure phase state of the form \cite{PLA03}
\begin{equation}
|f\rangle=\sum_{n\in
\mathcal{Z}_q}u_n|n\rangle,~~u_n=\frac{1}{\sqrt{q}}\exp(in
\beta),
\end{equation}
where $\beta$ is a real parameter, and we computed the phase
expectation value
$\langle \Theta_{\rm{lock}} \rangle=\sum_k\theta'_k|\langle \theta'_k|f \rangle|^2$ which
reads
\begin{equation}
\langle
\Theta_{\rm{lock}}\rangle=\frac{\pi}{q^2}\sum_{n,l}c_q(l-n)\exp(i\beta(n-l)).
\label{expec2}
\end{equation}
For $\beta=1$ it was found that $\langle
\Theta_{\rm{lock}}\rangle$ has more pronounced peaks  at
those values of $q$ which are precisely powers of a prime
number, and it can be approximated by the normalized Mangoldt function
$\pi \Lambda(q)/\ln q$. For $\beta=0$ the expectation value of
$\langle \Theta_{\rm{lock}}\rangle$ is much lower. The parameter
$\beta$ can be used to minimize the phase uncertainty well below
the classical value \cite{PLA03}. Related phase fluctuations, reflecting properties
of the distribution of prime numbers, were
obtained in the frame of a quantum statistical mechanics of shift
and clock operators. This algebra was also found relevant as a
model of time perception \cite{PlanatNeuro}.

Finally, the phase fluctuations arising from the quantum phase
states in MUBs are found to be related to Gaussian sums of the
form
\begin{equation}
G(\psi,\kappa)=\sum_{x \in F_q^*}\psi(x)\kappa(x). \label{Gauss}
\end{equation}
Using the notation $\psi_0$ for a trivial multiplicative character,
$\psi=1$, and $\kappa_0$ for a trivial additive character,
$\kappa=1$, Gaussian sums (\ref{Gauss}) satisfy
$G(\psi_0,\kappa_0)=q-1$; $G(\psi_0,\kappa)=-1$;
$G(\psi,\kappa_0)=0$ and $|G(\psi,\kappa)|=q^{1/2}$ for nontrivial
characters $\kappa$ and $\psi$. We need, however, a  more general
expression
\begin{equation}
G(\psi,\kappa)=\sum_{x \in F_q}\psi(f(x))\kappa(g(x)),
\label{Gaussnew}
\end{equation}
where $f,g \in F_q[x]$, which is found to be of the order of
magnitude $\sqrt{q}$ (\cite{Lidl83}, p.\,249). As a matter of fact,
the two factors in the expression for the probability distribution
$\langle \Theta_{\rm{Gal}}\rangle=\sum_{b \in F_q}\theta_b|\langle \theta_b|f \rangle |^2$
have absolute values bounded by the absolute value of generalized
Gauss sums (\ref{Gaussnew}), so that $|\langle \theta_b|f \rangle |^2 \le
\frac{1}{q}$ as it can be expected for an arbitrary phase factor.
To be more rigorous,  the phase expectation value can be expressed as
\begin{equation}
\langle \Theta_{\rm{Gal}}\rangle=\frac{2 \pi}{q^3}\sum_{m,n \in
F_q}\psi_k(m-n)\exp[i(n-m)\beta]\omega_p^{tr[a(m^2-n^2)]}S(m,n),
\label{expect}
\end{equation}
where  $S(m,n)$ was defined in Sect.\,3.3. All the $q$ diagonal
terms $m=n$ in $\langle\Theta_{\rm{Gal}}\rangle$ contribute an order of
magnitude $\frac{2\pi}{q^3}q S(n,n)\simeq \pi$. The contribution
of off-diagonal terms and possible cancellation of phase
oscillations in the phase expectation value and in phase variance
are discussed in \cite{Planat05}.

\section{Mutual Unbiasedness and Maximal Entanglement}

As we have shown, there is a founding link
between irreducible polynomials over a ground field $F_p$ and
complete sets of mutually unbiased bases arising from the Fourier
transform over a lifted field $F_q$, $q=p^m$.
On the other hand, the physical concept of entanglement over a
Hilbert space $\mathcal{H}_q$ evokes irreducibility. Roughly
speaking, entangled states in $\mathcal{H}_q$ cannot be factored
into tensorial products of states in Hilbert spaces of lower
dimension. We will now show that there is an intrinsic relation between
MUBs and maximal entanglement.

We are all familiar with the Bell states
\begin{eqnarray}
&(|\mathcal{B}_{0,0}\rangle,|\mathcal{B}_{0,1}\rangle)=\frac{1}{\sqrt{2}}(|00\rangle+|11\rangle,|00\rangle-|11\rangle),\nonumber\\
&(|\mathcal{B}_{1,0}\rangle,|\mathcal{B}_{1,1}\rangle)=\frac{1}{\sqrt{2}}(|01\rangle+|10\rangle,|01\rangle-|10\rangle),
\nonumber
\end{eqnarray}
where a compact notation $|00\rangle=|0\rangle\odot|0\rangle$,
$|01\rangle=|0\rangle\odot|1\rangle$, {\it etc.} is employed for the
tensorial products. These states are both orthonormal and maximally entangled, such
that $trace_2|\mathcal{B}_{h,k}\rangle \langle\mathcal{B}_{h,k}|
=\frac{1}{2}I_2$, where $trace_2$ is the partial trace over the
second qubit \cite{Nielsen00}.
One can define more generalized Bell states using the
multiplicative Fourier transform \cite{Pegg89} applied to the
tensorial products of two qudits \cite{Planat05}, viz.
\begin{equation}
|\mathcal{B}_{h,k}\rangle=\frac{1}{\sqrt{q}}\sum_{n=0}^{q-1}\omega_q^{k
n}|n,n+h\rangle. \label{FourierEntang}
\end{equation}
These states are both orthonormal, $\langle
\mathcal{B}_{h,k}|\mathcal{B}_{h',k'} \rangle
=\delta_{hh'}\delta_{kk'}$, and maximally entangled,
$trace_2|\mathcal{B}_{h,k}\rangle \langle\mathcal{B}_{h,k}|
=\frac{1}{q}I_q$.

For odd characteristic, we can also define a more general
class of maximally entangled states, using the Fourier transform
over $F_q$ and Eq. (\ref{MUB1}), as follows
\begin{equation}
|\mathcal{B}_{h,b}^a\rangle=\frac{1}{\sqrt{q}}\sum_{n=0}^{q-1}\omega_p^{tr[(a
 n + b)n ]}|n,n+ h\rangle.
\label{entangledGalois}
\end{equation}
A list of the generalized Bell states of qutrits for the base
$a=0$ can be found in \cite{Fujii01} , the work that relies on a
coherent state formulation of entanglement. In general, for $q$ a
power of a prime, starting from (\ref{entangledGalois}) one
obtains $q^2$ bases of $q$ maximally entangled states. Each set of
the $q$ bases (with $h$ fixed) has the property of mutual
unbiasedness.

\section{Mutually Unbiased Bases in Even Characteristic}
\label{Even}
\subsection{Construction of the Galois rings of characteristic four}

The Weil sums (\ref{Weilsums}), which have been proved useful
in construction of MUBs in odd characteristic, are not useful for
$p=2$ since in this case the degree $d$ of the polynomial $f(x)$ is such that
$\rm{gcd}$$(d,q)=2$ --- the characteristic of the relevant Galois fields.
An elegant method for constructing complete sets of MUBs of
$m$-qubits was found  in \cite{Klapp03}. It makes use of algebraic
objects in the context of quaternary codes \cite{Wan97}, the so-called
Galois rings $R_{4^m}$. In contrast to the Galois fields where the
ground alphabet has $p$ elements in the field
$F_p=\mathcal{Z}_p$, the ring $R_{4^m}$ takes its ground alphabet
in $\mathcal{Z}_4$. To construct this ring one uses the ideal class
$(h)$, where $h$ is a (monic) basic irreducible polynomial of
degree $m$ such that its restriction to $\bar{h}(x)=h(x)
~\rm{mod}~ 2$ is irreducible over $\mathcal{Z}_2$. The Galois ring
$R_{4^m}$ is the residue class ring
$\mathcal{Z}_4[x]/(h)$ and has cardinality $4^m$.

We also need the concept of a primitive polynomial. To this end, we recall
that a (monic) primitive polynomial, of degree $m$, in the ring
$F_q[x]$ is irreducible over $F_q$ and has a root $\alpha \in
F_{q^m}$ that generates the multiplicative group of $F_{q^m}$. A
polynomial $f\in F_q[x]$ of degree $m$ is primitive iff $f(0)\neq
0$ and divides $x^r-1$, where $r=q^m-1$.
Similarly for Galois rings $R_{4^m}$, if $\bar{h}[x]$ is a
primitive polynomial of degree $m$ in $\mathcal{Z}_2[x]$, then
there is a unique basic primitive polynomial $h(x)$ of degree $m$
in $\mathcal{Z}_4[x]$ (it divides $x^r-1$, with $r=2^m-1$). It can
be found as follows \cite{Planat05}. Let $\bar{h}(x)=e(x)-d(x)$,
where $e(x)$ contains only even powers and $d(x)$ only odd powers;
then $h(x^2)=\pm(e^2(x)-d^2(x))$. For $m=2$, $3$ and $4$ one takes
$\bar{h}(x)=x^2+x+1$, $\bar{h}(x)=x^3+x+1$ and
$\bar{h}(x)=x^4+x+1$ and gets $h(x)=x^2+x+1$, $x^3+2x^2+x-1$
and $x^4+2x^2-x+1$, respectively.

Any non zero element of $F_{p^m}$ can be expressed in
terms of a single primitive element. This is no longer true in
$R_{4^m}$, which contains zero divisors. But in the latter case
there exists a nonzero element $\xi$ of order $2^m-1$ which is a
root of the basic primitive polynomial $h(x)$. Any element $y \in
R_{4^m}$ can be uniquely determined in the form $y=a + 2 b$, where
$a$ and $b$ belong to the so-called Teichm\"{u}ller set
$\mathcal{T}_m = (0,1,\xi,\cdots,\xi^{2^m-2})$. Moreover, one
finds that $a=y^{2^m}$. We can also define the generalized trace
to the base ring $\mathcal{Z}_4$ as the map
\begin{equation}
\widetilde{\rm{tr}}(y)=\sum_{k=0}^{m-1}\sigma^k(y), \label{trace2}
\end{equation}
where $\sigma(y)$  is the so-called Frobenius
automorphism, obeying the rule
\begin{equation}
\sigma(a+2 b)=a^2+ 2 b^2.
\end{equation}
In $R_{4^m}$ the additive characters
acquire the form
\begin{equation}
\tilde{\kappa}(x)=\omega_4^{\widetilde{\rm{tr}}(x)}=i^{\widetilde{\rm{tr}}(x)}.
\end{equation}
\subsection{Exponential sums over $R_{4^m}$}

The Weil sums (\ref{Weilsums}) are replaced by the following exponential
sums \cite{Klapp03}
\begin{equation}
\Gamma(y)=\sum_{u \in \mathcal{T}_m}\tilde{\kappa}(y u),~~y \in
R_{4^m}, \label{newWeilsums}
\end{equation}
which satisfy
\begin{equation}
|\Gamma(y)|=\left\{\begin{array}{ll}
&0~~~~~~~~\mbox{if}~y \in 2 \mathcal{T}_m,~y \ne 0, \\
&2^m~~~~~~\mbox{if}~y=0,\\
&\sqrt{2^m}~~~~\mbox{otherwise}.
\end{array}\right.
\end{equation}
Gauss sums for Galois rings were constructed in \cite{Oh01} and are of the form
\begin{equation}
G_y(\tilde{\psi},\tilde{\kappa})=\sum_{x \in R_{4^m}}
\tilde{\psi}(x)\tilde{\kappa}(yx), ~~y \in
R_{4^m},\label{GaussGal}
\end{equation}
where the multiplicative character $\bar{\psi}(x)$ can be made
explicit.
Using the notation $\tilde{\psi_0}$ for a trivial multiplicative
character and $\tilde{\kappa_0}$ for a trivial additive character,
we get
$G(\tilde{\psi_0},\tilde{\kappa_0})=4^m$,
$G(\tilde{\psi},\tilde{\kappa_0})=0$ and
$|G(\tilde{\psi},\tilde{\kappa})|\le 2^m$.

\subsection{Mutually unbiased bases of $m$-qubits} \label{sect:MUB2s}

It was mentioned in the previous section that each element $y$ of the
ring $R_{4^m}$ can be decomposed as $y=a+2b$, with $a$ and $b$ belonging to the
Teichm\"{u}ller set $\mathcal{T}_m$. Employing this fact in the
character function $\tilde{\kappa}$, one obtains
\begin{equation}
|\theta_b^a\rangle=\frac{1}{\sqrt{2^m}}\sum_{n\in
\mathcal{T}_m}\tilde{\psi}_k(n)\tilde{\kappa}[(a+2b)n]|n\rangle,~~a,b
\in \mathcal{T}_m. \label{MUB1pair}
\end{equation}
This defines a set of $2^m$ bases (with index $a$) of $2^m$
vectors (with index $b$). Using Eq.\,(\ref{newWeilsums}), it is easy to show that the bases are
orthogonal and mutually unbiased to each other and to the
computational base. (For the explicit derivation of the bases, see
\cite{Planat05}.)

Quantum phase states of $m$-qubits (\ref{MUB1pair}) derive from a
``Galois ring" quantum phase operator as in Eq.\,(\ref{GalMUB}),
and calculations similar to those in Sect.\,\ref{MUBodd}
can be performed, since the trace operator defined by Eq.\,(\ref{trace2})
obeys the rules similar to those of the field trace operator,
Eq.\,(\ref{trace0}). In analogy to the case of qudits in dimension
$p^m$, $p$ an odd prime, the phase properties for sets of $m$-qubits
rely substantially on Eq.\,(\ref{GaussGal}). As before, the
calculations are tedious, but they can successfully be accomplished in
specific cases.

\section{Quantum Geometry From Projective Planes}

We have related complete sets of MUBs in dimension $p^m, ~p$
odd, to
additive characters over a Galois field. Complete sets of MUBs
offer an intriguing geometrical interpretation, being related to
discrete phase spaces \cite{Wootters04bis}, finite projective
planes \cite{Saniga, Saniga2}, convex polytopes
\cite{Bengtson}, and complex projective $2$-designs \cite{Klap2}.
The last-mentioned paper also points out an interesting link to symmetric
informationally complete positive operator measures (SIC-POVMs)
\cite{Grassl04} and to Latin squares \cite{Wocjan04}.
We focus here on the relation of MUBs to finite geometries and
projective planes.

\subsection{Mutually unbiased bases and projective planes}

A remarkable link between mutually unbiased measurements
and finite projective geometry has recently been noticed
\cite{Saniga}. Let us find the minimum number of different
measurements we need to determine uniquely the state of an
ensemble of identical $q$-state particles. The density matrix of
such an ensemble, being Hermitean and of unit trace, is specified
by $(2q^2/2)-1=q^2-1$ real parameters. When one performs a
non-degenerate orthogonal measurement on each of many copies of
such a system one eventually obtains $q-1$ real numbers (the
probabilities of all but one of the $q$ possible outcomes). The
minimum number of different measurements needed to determine the
state uniquely is thus $(q^2-1)/(q-1)= q+1$
\cite{Wootters89,Wootters04bis}.

\begin{figure}[t]
%\vspace*{13pt}
\centerline{\epsfig{file=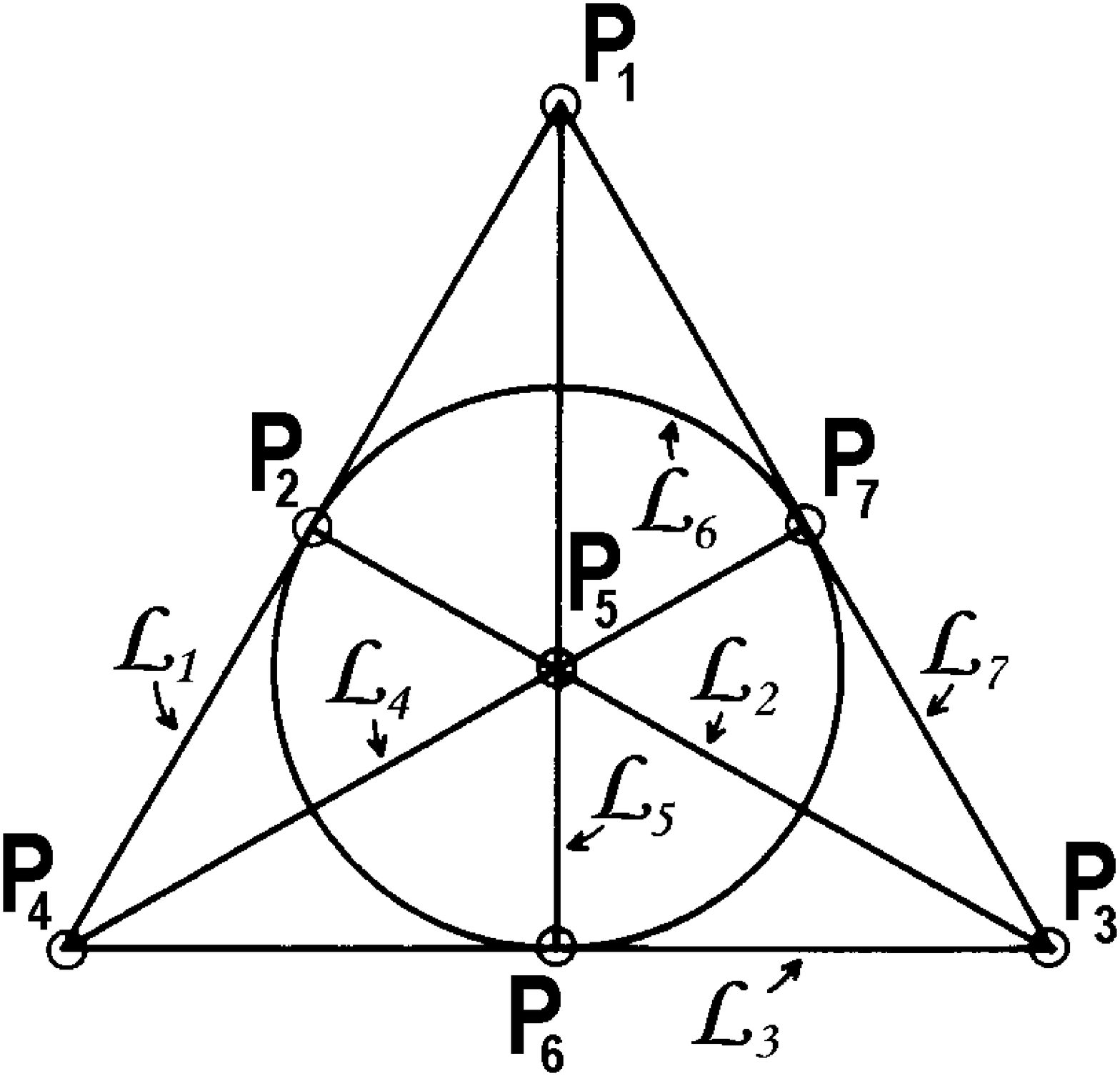, width=4.5cm}} %100 percent
\vspace*{12pt} \fcaption{\label{motion}The Fano plane; small circles (denoted as
$P_1$,\ldots,$P_7$) represent its points, while line-segments (${\cal L}_1$,\ldots,${\cal L}_5$ and
${\cal L}_7$) and a circle (${\cal L}_6$) stand for its lines. }
\end{figure}

It is striking that the identical expression can be found within
the context of finite projective geometry. A finite projective
plane is an incidence structure consisting of points and lines
such that any two points lie on just one line, any two lines pass
through just one point, and there exist four points, no three of
them on a line \cite{Beutel98}. From these properties it readily
follows that for any finite projective plane there exists an
integer $q$ with the properties that any line contains exactly
$q+1$ points, any point is the meet of exactly $q+1$ lines, and
the number of points is the same as the number of lines, namely
$q^{2}+q+1$. This integer $q$ is called the order of the
projective plane. The most striking issue here is that the order
of known finite projective planes is a power of prime. The
question of which other integers occur as orders of finite
projective planes remains one of the most challenging problems of
contemporary mathematics. The only ``no-go" theorem known so far
in this respect is the Bruck-Ryser theorem \cite{Beutel98} saying
that there is no projective plane of order $q$ if $q-1$ or $q-2$
is divisible by 4 and $q$ is not the sum of two squares. Out of
the first few non-prime-power numbers, this theorem rules out
finite projective planes of order 6, 14, 21, 22, 30 and 33.
Moreover, using massive computer calculations, it was proved that
there is no projective plane of order ten. It is surmised that the
order of any projective plane is a power of a prime.

It is conjectured \cite{Saniga} that the question of the existence
of a set of $q+1$ mutually unbiased bases in a $q$-dimensional
Hilbert space if $q$ differs from a power of a prime number is
identical with the problem of whether there exist projective
planes whose order $q$ is not a power of a prime number.

The smallest projective plane, also called the Fano plane (see Fig.\,1), is
obviously the $q=2$ one; it contains $7$ points and $7$ lines, any
line contains $3$ points and each point is on $3$ lines.
It may be viewed as a $3$-dimensional vector space over the field
$GF(2)$, each point being a triple $(g_1,g_2,g_3)$, excluding the
(0,0,0) one, where $g_i \in GF(2) = \{0,1\}$ \cite{Beutel98}. The
points of this plane can also be represented in terms of the
non-zero elements of the Galois field $G=GF(2^3)$, see the last column of Table\,1.

\subsection{Cyclic codes and projective spaces}

As shown in Sect.\,\ref{codes}, a linear code $C$ is a subspace of $F_q^n$,
$q=p^m$. And a  cyclic code is an ideal $(g)$ in the polynomial
field $F_q[x]\equiv F_q^n$ attached to a polynomial $g$
irreducible over $F_q$. One defines the Hamming distance
\cite{Lidl83} between $x$ and $y$ in $F_q^n$ as the number of
coordinates in which $x$ and $y$ differ. The minimum distance of a
code is an important concept which characterizes its efficiency
for error correcting;  it is defined as
\begin{equation}
d=d_{min}(C)=\min_{\stackrel{u,v \in C}{u \neq v }}d(u,v).
\end{equation}
A linear code corrects up to $[\frac{d-1}{2}]$ and detect up to
$d-1$ errors. It can be shown that for a linear $[n,k]$ code, the
following bound holds
\begin{equation}
d \le n-k+1=d_{\rm{max}}.
\end{equation}
A minimum distance code (or a maximum distance separable, MDS code)
is such that $d=d_{\rm{max}}$ and it is usually referred to as a $[n,k,d]$ code (or
$[n,n-r,r+1]$ code).
The binary Hamming $[7,4]$ code introduced in Sect.\,\ref{codes} thus
corrects up to $1$ and detect up to $3$ errors. It is the MDS
$[7,4,4]$ code.

There exists an intimate link between this code and the Fano plane,
which can be inferred as follows. Let us take its seven codewords $1$ to $7$
by cyclic extending of matrix (5), viz.
\begin{equation}
 \left [\begin{array}{ccccccc} 1 & 1&0&1&0&0&0\\ 0 & 1&1&0&1&0&0\\
 0&0&1&1&0&1&0\\0&0&0&1&1&0&1\nonumber \\1&0&0&0&1&1&0\nonumber \\
 0&1&0&0&0&1&1\nonumber \\1&0&1&0&0&0&1\nonumber \\
 \end{array}\right].
%\equiv
% \left [\begin{array}{c} P1 \\P2\\
%P3\\P4\\P5 \\
%P6\\P7 \\
% \end{array}\right].
 \label{table2}
\end{equation}
The above matrix is nothing but the incidence matrix of the Fano plane, obtained as follows: if $P_j$ is the $j\rm{th}$ point and
${\cal L}_i$ represents the $i\rm{th}$ line of the Fano plane, the elements of the matrix are
\begin{equation}
a_{ij}= \left\{\begin{array}{ll}
1 &~~\mbox{if}~ P_j \in {\cal L}_i,\\
 0 &~~ \mbox{otherwise}.\\
\end{array}\right.
\end{equation}

The link between good codes and projective geometry has recently received considerable
attention \cite{Beutel98}. Let us define a vector space $V$ of
dimension $\delta+1\ge 3$ over $F_q$. Then a projective geometry $P(V)$
can be defined as follows. The points of $P(V)$ are its $1$-dimensional
subspaces, the lines its $2$-dimensional subspaces and the
incidence structure in $P(V)$ is the set-theoretical containment. The
geometry $P(V)$ is the projective space coordinatized by the
Galois field $F_q$. This $\delta$-dimensional projective space $P(V)$ over
$F_q$ is usually denoted as $PG(\delta,q)$.

Next, a set of points in $P=PG(\delta,q)$ is called an arc if any $\delta+1$ of its
points form a basis of $P$. An arc having $n$ points is called an
$n$-arc.
In a projective plane $PG(2,q)$ an $n$-arc is a set of $n$ points
no three of which are collinear. If each point of an $n$-arc is
exactly on one tangent, the arc is called an oval. The maximum value of
$n$ for an $n$-arc is
\begin{equation}
m(2,q)=\left\{\begin{array}{ll}
&q+1~~\mbox{when}~q~\mbox{is~odd}, \nonumber\\
&q+2~~\mbox{when}~q~\mbox{is~even}. \nonumber\\
\end{array}\right.
\label{(ovals)}
\end{equation}
The meaning of Eq.\,(\ref{(ovals)}) is as follows. If $q$ is odd, then
arcs with a maximum number of points are  ovals. If $q$ is
even then, each oval can be uniquely extended to a $(q+2)$-arc,
which is called a hyperoval. The possible correspondence between
ovals and complete set of MUBs is discussed in \cite{Saniga2}.

There is a one to one correspondence between the generator matrix
of [$n,n-r$] MDS codes and $n$-arcs in $PG(r-1,q)$
(\cite{Hirschfeld98}, p.\,73). The construction of good codes with a
prescribed minimum distance can be rephrased as follows. One is
given the minimum distance $d$ and $r$. Determine the greatest
length of the code, $\max_{d-1}(r,q)$.

The simplest case is $d=3$ for which $\max_2(r,q)$ is the maximum
possible number of points in $PG(r-1,q)$ such that two of them are
independent: this is, obviously, the total number of points in $PG(r-1,q)$, and
thus
\begin{equation}
{\rm max}_{2}(r,q)=q^{r-1}+\cdots+q+1.
\end{equation}
In particular, we have $\max_{2}(r,2)=2^r-1$, which corresponds to the
Hamming $[n,n-r]$ code.
The case $d=4$ is less trivial. Only partial results are known:
\begin{equation}
{\rm max}_3(r,2)=2^{r-1}, \label{(a)}
\end{equation}
\begin{equation}
{\rm max}_3(3,q)=\left\{\begin{array}{ll}
&q+1~~\mbox{when}~q~\mbox{is~odd}, \nonumber\\
&q+2~~\mbox{when}~q~\mbox{is~even}, \nonumber\\
\end{array}\right.\label{(b)}
\end{equation}
and
\begin{equation}
{\rm max}_3(4,q)=q^2+1. \label{(c)}
\end{equation}
Putting $r=3$ in Eq.\,(\ref{(a)}) one gets the
case of the $[7,4,4]$ code considered above. The geometry of
Eq.\,(\ref{(b)}) answers to ovals ($q$ odd) and hyperovals ($q$
even) of $PG(2,q)$, that of Eq.\,(\ref{(c)}) to
ovoids\footnote{An ovoid is a nonempty set $\mathcal{O}$ of
points of $PG(d,q)$ such that no three points of
$\mathcal{O}$ are collinear and for each point $Q \in
\mathcal{O}$ the tangents through $Q$ span the whole hyperplane.}
~of $PG(3,q)$.

\section{Conclusion}
Let us briefly summarize basic ideas developed in the paper.
At the algebraic level, there exists a polynomial field $F_q^n$ defined over a ``base" field $F_q$, with  $q=p^m$ and $p$ a prime.
The cyclic encoding (in \lq\lq time" $n$) comes from (cyclotomic) laws of  partitioning $F_q^n$.
At the geometrical level, a projective space can always be partioned into subsets (called spreads). For example, $PG(3,q)$ is
partioned into subsets of $q^2$+1 mutually skew (i.e. pairwise disjoint) lines. These may well form ovoids, which
``give rise" to $MDS-$codes. Similarly, there is a
partitioning of a projective plane $PG(2,q)$ into sets of $q$+1 lines. These may well be (the tangents of) ovals, which are conjectured
to reproduce properties of the sets of mutually unbiased bases. Here, we play with a two-dimensional  \lq\lq quantum space-time."
The vectorial projective space, $PG(\delta,q)$, is thus a promising playground for tackling both the measurement and coding problems
in quantum mechanics. Yet, there exist more general kinds of finite (projective) geometries, e.g. (non-)Desarguesian projective
planes defined over quasi-/near-fields, the latter obeying less
stringent rules than fields \cite{hp}. These are, we believe, candidates for addressing another intriguing quantum effects like partial
entanglement and decoherence.

\nonumsection{Acknowledgement}
\vspace*{-.2cm}
\noindent
M.S. would like to acknowledge the support received from a 2004
SSHN Physics Fellowship
%of the French Ministry of Youth, National Education and Research
($\#$411867G/P392152B).

\nonumsection{References}

\end{document}